\documentclass[12pt,preprint]{aastex}
\usepackage{graphics}

\newcommand{\um}{\mbox{{\usefont{U}{eur}{m}{n}\symbol{22}}m}}

\slugcomment{Submitted to ApJ: ; accepted: }
\shorttitle{}
\shortauthors{Kranz, Slyz, \& Rix}

\received{2002 July 26}
\begin{document}

\title{Dark matter within high surface brightness spiral galaxies}
\author{Thilo Kranz\altaffilmark{1}$^,$\altaffilmark{3}, Adrianne Slyz\altaffilmark{2} and Hans-Walter Rix\altaffilmark{1}}
\altaffiltext{1}{Max-Planck-Institut f\"ur Astronomie, K\"onigstuhl 17, 69117 Heidelberg, Germany;
 kranz@mpia.de, rix@mpia.de}
\altaffiltext{2}{Nuclear and Astrophysics Laboratory, Keble Road, OX1 3RH Oxford, UK; slyz@astro.ox.ac.uk}
\altaffiltext{3}{now at: German Aerospace Center DLR, K\"onigswinterer Stra{\ss}e 522 -- 524, 53227 Bonn, Germany;
 thilo.kranz@dlr.de}

\begin{abstract}
We present results from a detailed dynamical analysis of five high surface brightness, late type
spiral galaxies NGC 3810, NGC 3893, NGC 4254, NGC 5676, and NGC 6643
which were studied with the aim to quantify the luminous-to-dark matter
ratio inside their optical radii. The galaxies' stellar light distribution and gas kinematics have
been observed and compared to hydrodynamic gas simulations which predict the
gas dynamics arising in response to empirical gravitational potentials, which are combinations of differing stellar
disk and dark halo contributions. The gravitational potential of the stellar disk was derived from
near-infrared photometry, color-corrected to yield a constant stellar mass-to-light ratio (M/L); for the dark halo,
the mass density distribution of an axisymmetric isothermal sphere with a core was chosen. Hydrodynamic
gas simulations were performed for each galaxy for a sequence of five different mass fractions of the stellar disk and
for a wide range of spiral pattern speeds. These two parameters mainly determine the modelled gas
distribution and kinematics.
The agreement between the simulated and observed gas kinematics permitted us to conclude that
the galaxies with the highest rotation velocities tend to possess very massive stellar disks which dominate the gas
dynamics within the optical radius. In less massive galaxies, with a maximal rotation velocity of
$<$\,200\,km\,s$^{-1}$, the mass of the dark halo at least equals the stellar mass within 2 -- 3 disk scale lengths.
The maximal disk stellar
mass-to-light ratio in the $K$-band was found to lie at about M/L$_K \approx 0.6$.
Furthermore, the gas dynamic simulations provide a powerful tool to accurately determine the dominant
spiral pattern speed for galaxies, independent of a specific density wave theory. It was found that the location of
the corotation resonance falls into a narrow range of around three exponential disk scale lengths for all galaxies
from the sample. The corotation resonance encloses the strong part of the stellar spiral in all cases.
Based on the experience gained from this project, the use of a color-correction to account for local stellar
population differences is strongly encouraged when properties of galactic disks are studied that rely on
their stellar mass distributions.
\end{abstract}


\section{INTRODUCTION}

Much evidence has accumulated in recent years that the stellar disks of high
surface brightness spiral galaxies dominate the mass budget of the inner regions. Most of the studies
arguing for a maximal disk scenario, are based on the detailed analysis of high resolution rotation
curve measurements \citep{bla99,pal00,rat00,sal01}.
These studies, however, generally derive the rotational support of the stellar disk from an axisymmetric
disk mass model and allow no consideration of non-circular rotation components. More sophisticated
modelling strategies have been applied to spiral galaxies with variable significance \citep{eri99,pig01}.
For the most part those studies also support heavy disks for high surface brightness galaxies,
but they also find candidates for which lower disk mass fractions are more likely.
The most convincing direct arguments for the maximal disk scenario come from the studies of strong bars
in spiral galaxies. These features induce a strong dynamic trace in the velocity field and provide a
good laboratory for estimating the stellar mass component. Based on fluid dynamic modelling, a maximal
disk solution is found by \citet{eng99} for the Milky Way and by \citet{wei01}
for NGC 4123. Furthermore, theoretical considerations hint for the requirement of a
non-massive halo contribution in the central regions of strongly barred galaxies with a very cold disk,
because otherwise dynamical friction would slow down the bar very quickly, leading to its destruction
\citet{deb98,deb00}. On the other hand, for initially hotter disks, this effect is largely reduced
\citet{ath02} and references therein.

However, there is also evidence that even high surface brightness spiral galaxies might be dominated
by the dark matter mass component in their central disk regions as suggested by high resolution simulations
of cosmological dark matter halos \citep[e.g.]{moo99a,ghi00,fuk01}. \citet{bot93} inferred from stellar velocity
dispersions in spiral galaxy disks that a more massive halo component is needed to explain the findings.
There are recent studies that argue for lighter disk models by making use of previously hardly exploited
relations. \citet{CR99} applied a statistical Tully-Fisher relation analysis to a large
sample of galaxies trying to relate the maximal rotation velocity of a galaxy to its disk size. \citet{mal00},
and more recently \citet{tro02} used the geometry of gravitationally lensed systems to disentangle the
effects of the stellar disk and the halo masses.
These groups found that the dark halo also dynamically plays an important role in the galaxies' inner regions.

In the present study, we quantify the relative mass fractions of the stellar disk and the inner dark
halo by analyzing the non-axisymmetric force component in the gravitational force field that arises in
response to the spiral structure
of the observable stellar mass distribution. We address this issue by exploring the influence of the spiral arms
on the gas velocity field that builds up in response to the total gravitational potential of the galaxy. We
model the gravitational potential from
a mass map of the stellar distribution and a dark halo model. With the use of 
hydrodynamical gas simulations, we predict the gas velocity fields for different empirical mass models
of the sample galaxies, assembled from a variety of stellar disk and dark matter mass components.
The two parameters that dominate the resulting forces in the disk are the stellar disk mass contribution
and the pattern speed of the stellar spiral, $\Omega_{\rm p}$. Thus, in this study we explore the effects of
those parameters and compare the simulated gas morphology and velocity wiggles
with our observations, enabling the identification of the best fitting scenario for each individual galaxy.
Comparable studies have been applied to barred spiral
galaxies taking advantage of the strong dynamical trace of the bar in the velocity field \citep{wei01}.

First results from our study have been presented already in \citet{kra01}, (hereafter, KSR01) for NGC 4254 where
it was proven that this strategy yields valuable information which allows us to constrain disk
and halo fractions also for spiral galaxies without a strong bar. In the present paper we only briefly discuss the
modelling and comparison strategies involved (Section~\ref{modelling}) but instead focus on the conclusions
of the analysis. More details about the modelling issues can be found in KSR01, \citet{sly02b}
and \citet{kra02}.

\section{OBSERVING AND MODELLING THE SAMPLE GALAXIES} \label{modelling}

For the studied galaxies we collected near-infrared (NIR) photometry as the basis to study the
luminous mass distribution and gas emission line (H$\alpha$) spectroscopy to acquire kinematic
information for the systems.
The NIR images were taken during two observing runs in May 1999 and March 2000 with the Omega
Prime camera at the Calar Alto 3.5 m's prime focus \citep{biz98}. It provides a field
of view of 6\farcm76 $\times$ 6\farcm76 with a resolution of 0\farcs3961 per pixel. We used
the K\arcmin filter which has a central wavelength of 2.12 \um. 
We obtained the gas kinematics from longslit spectroscopy measurements of the H$\alpha$ emission.
With the setup we used, the TWIN spectrograph that is located at the cassegrain focus of the
Calar Alto 3.5 m telescope achieved a spectral resolution of 0.54 \AA~per
detector pixel, which translates to $24.8\,{\rm km\,s}^{-1}$ LOS-velocity resolution per pixel,
allowing us to determine LOS-velocities with $\sim 7\,{\rm km\,s}^{-1}$ precision. We sampled the
velocity field of the galaxy along 14 or 16 slit position angles. The
slit of the TWIN spectrograph measures 4\arcmin $\times$ 1\farcs5 on the sky.
The spectra were obtained during three observing runs in June 1999, May and December 2000. 

The modelling procedure involved four separate steps. First, a stellar mass map had to be derived
for each galaxy from the NIR images, as the basis to calculate the gravitational disk potential.
Then, a parameterized dark matter component had to be added to the stellar mass component,
constrained to render the observed rotation curve of the galaxy. In a third step we carried out
two-dimensional hydrodynamical gas simulations in these gravitational potentials, modelling the gas
density distribution and the gas velocity field. Finally, these models needed to be compared to the
observations to draw conclusions about the galaxies' structure. This detailed analysis was performed
for the four galaxies \objectname{NGC 3810}, \objectname{NGC 3893}, \objectname{NGC 5676} and
\objectname{NGC 6643}, following KSR01. We included \objectname{NGC 4254} in the final analysis, which
was published in KSR01.

\subsection{Constructing the stellar mass maps} \label{coco}

For the first step, converting the stellar light map into a stellar surface mass map of unknown stellar M/L,
we used color-corrected NIR photometry, applying a correction recipe provided by \citet{bel01}. Bell \& de Jong showed,
based on population synthesis models and the assumption of a universal initial mass function, that a galaxy's
color is tightly correlated to its mass-to-light ratio. This applies to a galaxy's global color as well as to local
colors, and is approximately true, regardless of whether the color changes are due to dust, age, or metallicity
differences. The color--M/L relation can be employed to improve the estimate of the stellar surface mass
density distribution, $\Sigma_{\star}$, of galaxies taken from the measured NIR surface brightness, $\mu_K$,
by using color maps of the galaxy:
\begin{equation} 
2.5 \log \Sigma_{\star} = - \mu_K + a_K (V - K) + c_K \label{equation}
\end{equation} 
for the example of $(V - K)$ color maps. The optical images that were required to determine the galaxy colors
were taken from the literature (see Table~\ref{properties}).
Although this correction is physically motivated, it relies on empirical constants $a$ and $c$\footnote{The
constant $c$ combines both empirical parameters and ones that are specific to the particular galaxy that
is studied: $c = c\prime + \mu_{\odot}$, where $c\prime$ is the empirical constant and $\mu_{\odot}$ is 
the surface brightness of 1\,M$_{\odot}$ at the galaxy's physical distance.} determined from numerical studies.
Following Bell \& de Jong (2001, Table~1), we used $a_K = 0.314$ and values for $c_K$, depending on the
galaxies' physical distances.

\begin{figure}
 \resizebox{\hsize}{!}{\includegraphics{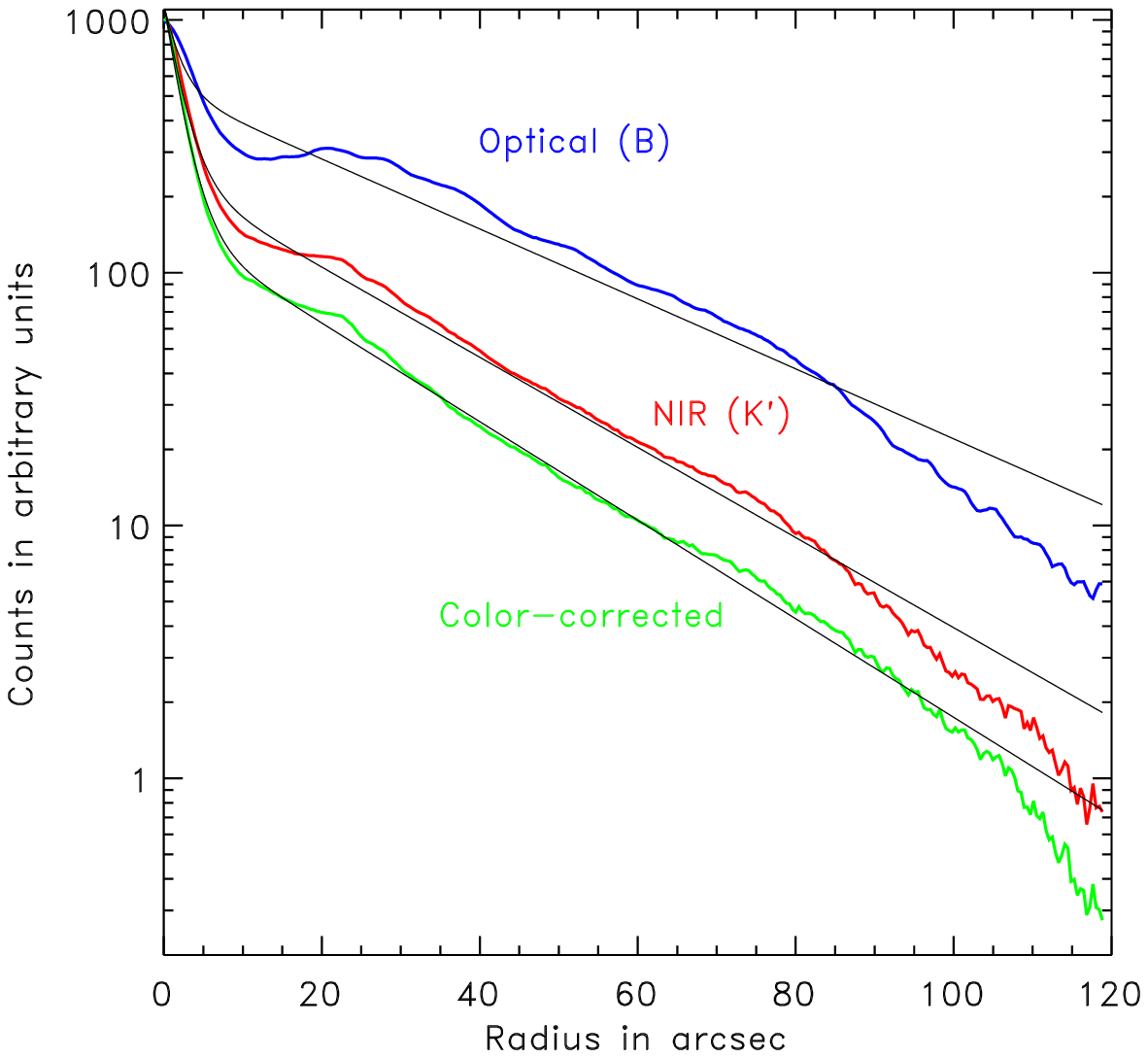}}
 \caption{The effects of the color correction, displayed for the example of NGC 6643.
Displayed are the B-band, K-band and color corrected profiles. Due to the outward blueing of
the disk, the color corrected profile becomes steeper than the other two profiles. The smooth lines
are fits with double exponential profiles. \label{n6643prof}}
\end{figure}

When applied to our sample galaxies, this color correction (Eqn.~\ref{equation}) had two major consequences.
First, the radial disk scale length shortened from the K-band image to the color corrected stellar mass image.
Second, the detailed two dimensional distribution changed by attenuating or enhancing different azimuthal
light features in the disk. Figure~\ref{n6643prof} illustrates how the color correction affects the radial
light distribution of NGC 6643. The steepening of the radial profile is due to the overall color gradients within the
disk. Interestingly, the color correction also diminishes the deviation of the stellar profile from an exponential
disk, which in the stellar light appear in part caused by different stellar populations.
Despite the color correction, the derived mass map still may bear some systematic
uncertainties for NGC 5676 and NGC 6643, because these galaxies are classified as starburst galaxies.
The correction (Eqn.~\ref{equation}) need not be a good approximation in this regime.

The color correction not only applies to population differences, but also works for
moderate dust absorption ($\tau_{\rm opt} < 1$). Dust makes the stellar radiation dimmer and redder. To
first order, this is accounted for by Eqn.~\ref{equation}, with similar constants. 
Eventually, even in the $K$-band, which is usually considered as quite a good estimate of the stellar
mass distribution, the color correction changes the radial scale length of a galaxy by about 10\,\%.
Consequently, this causes the inferred radial force components in the disk to change their magnitude,
affecting the shape of the rotation curve. The use of more than only two colors for the correction, could
further improve the quality of the mass density distribution that can be derived.
Besides this color correction, several other minor procedures (e.g., spatial filtering) have been
applied to the $K$-band images.
They were described in more detail in KSR01 and \citet{kra02}.

\subsection{Gravitational potentials and hydrodynamic models}

In order to derive total gravitational potentials, $\Phi_{\rm tot}$, for the galaxies, a maximal stellar
M/L was chosen for the disk, which corresponds to a maximal stellar potential contribution, $\Phi_{\star}$.
For five different values of the stellar M/L, parameterized by a disk fraction ${\rm f_d}$,
the total gravitational potential of the galaxy was assembled by adding the potential of a dark halo
component, constrained to match the observed rotation curve of the galaxy. For the dark halo, the mass
density distribution of
an axisymmetric isothermal sphere with a core was chosen. Its gravitational potential, $\Phi_{\rm halo}$, was
also calculated in the two-dimensional plane of each galaxy's disk:
\begin{equation} \label{eqn2}
\Phi_{\rm tot}(\mbox{\boldmath$R$}|{\rm f_d}) = {\rm f_d}\, \Phi_{\star}(\mbox{\boldmath$R$}) + \Phi_{\rm halo}(\mbox{\boldmath$R$}|{\rm f_d})
\end{equation}
with ${\rm f_d} = 0.2, 0.45, 0.6, 0.85$ and 1. Finally, a fixed pattern speed was chosen to represent the dominant
spiral feature, leading to centrifugal and coriolis forces in the simulations.

The hydrodynamic simulations were performed using the Bhatnagar-Gross-Krook (BGK) hydro-code,
of Adrianne Slyz. This is an Eulerian, grid based hydrodynamics code which is derived
from gas kinetic theory. Its concept and performance is well documented in \citet{xu98}, \citet{sly98},
\citet{sly99},  \citet{sly02a}. The gas density distributions and gas velocity fields
resulting from these simulations were compared to the observed galaxy morphologies and to the measured
\ion{H}{2} kinematics (see Kranz 2002 for modelling and comparison details). The evaluation of the data match
let us probe both the relative luminous-to-dark matter
fractions for the sample galaxies and the pattern speed of the spiral structure.

\section{RESULTS FROM THE ANALYSIS}

For the discussion of the results we also include the findings for the spiral galaxy NGC 4254 which was
studied with the same approach and has been presented in KSR01. In light of this, our sample
consists of five high surface brightness spiral galaxies for which we discuss the modelling results in a 
common context.

\subsection{Disk dynamics}

The secular evolution of spiral structures in disk galaxies is not yet fully
understood. The debate is about whether spirals arise from quasi-stationary,
self-sustained density waves, or if they are transient features, triggered by internal or
external processes, that fade away unless continuously replenished \citep[for a review]{ath84}.
The present study is not very dependent on the long-term evolution of the spiral. The gravitational
potentials used for the modelling are calculated from the morphological ``snapshot'' that the
measurements provide and the potentials were not allowed to evolve during the simulation process.
To achieve a steady state flow in the simulations we needed
to adopt a spatially and temporally constant pattern speed for the dominant spiral structure.
The good qualitative match of observed and simulated morphology in our sample suggests that in the end this
approximation might not be far off. Furthermore, it must be noted that the modelling
only requires that the spiral morphology is stable over about one dynamical period, the time required
for the gas to adjust to moderate changes in the pattern speed. Aside from some subtle
evidence that weak central bars of our sample galaxies reveal pattern speeds that are decoupled from the rest of the
disk, the rigid spiral pattern assumption seems to be consistent for most of the disks studied here.
To avoid confusion with possible effects of these weak central bars, we exclude in such cases the very inner parts
of the disk from the comparison with the simulations and thus restrict ourselves to the galaxies' outer disks. 

\subsubsection{Location of the corotation resonance}

If a fixed pattern is a reasonable approximation, the determination of resonance locations within galactic
disks is well defined. It has already been proven
that such an approach yields valuable information about the disk dynamics of several galaxies, e.g., M 51
\citet{gar93}, M 100 \citet{gar94}, NGC 7479 \citep{sem95} and
M 94 and NGC 3310 \citep{mul96}.  However, since these studies were mostly dealing with single
systems, the results were not put into a more general context.

\begin{figure}
\centering
 \resizebox{14cm}{!}{\includegraphics{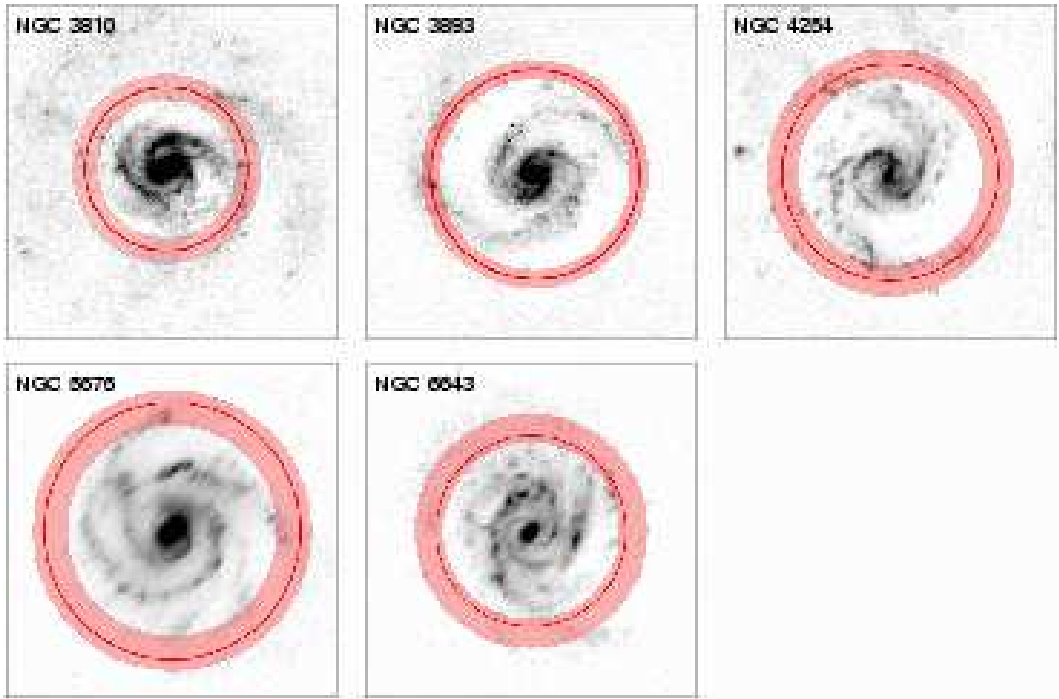}}
 \caption{Location of the corotation resonance, $R_{\rm CR}$, including its uncertainty range,
 overlaid onto deprojected, spiral enhanced $K$-band galaxy images. The $R_{\rm CR}$'s were derived from comparing the
 spiral arm morphology with the gas morphology in hydrodynamical simulations of differing pattern speeds.
 The values for $R_{\rm CR}$ are listed in Table~\ref{corotations}. Note that the corotation resonance
 appears to lie just outside the strong spiral. \label{allrccomp}}
\end{figure}

\begin{figure}
\centering
 \resizebox{14cm}{!}{\includegraphics{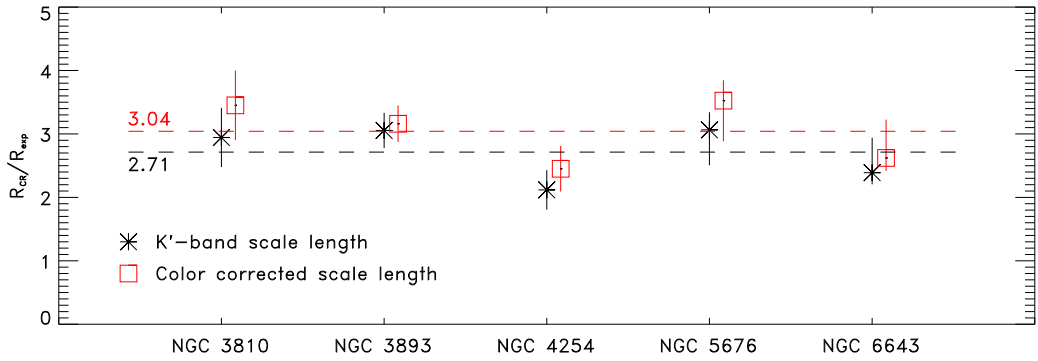}}
 \caption{Location of the corotation radius in terms of the exponential disk
scale length for the sample galaxies. The asterisks mark the values for the $K$-band
disk, while the boxes refer to the color corrected scale lengths.
The error bars come from uncertainties in the corotation radius estimations.
The horizontal lines represent mean values; the lower one for the $K$-band disk,
the upper one for the color corrected disk.\label{diskrcrexp}}
\end{figure}

KSR01 showed that it is possible with our modelling approach to locate the corotation resonances with a
fairly high precision, by comparing the
simulated gas density distributions for different spiral pattern speeds to the observed disk morphology.
The decision for the best pattern speed, or equivalently the corotation radius, $R_{\rm CR}$, was based on a
visual match in the morphologies, which accounted for the phase pitch angle and the extent of the spiral arms,
as well as for bifurcations. This method appears to be of a qualitative nature, and it might be argued that
the pitch angle of the response depends on the velocity dispersion of the gas, in the sense that cold material will
form a more tightly wound spiral than the stars. However, comparing the simulated gas response to images
of the galaxies taken in optical wavelengths rather than the NIR images, it can be seen that the most dense,
shocked regions in the simulations come to lie in fact inside the stellar spirals and tend to coincide with
dust lanes, in cases there are some. Furthermore, this approach was supported by the fact that the simulated
velocity field for the selected $R_{\rm CR}$ also matches best with the observed rotation curves and less good
for other pattern speed assumtions. In this way it is also possible to estimate an uncertainty range. Details
of the $R_{\rm CR}$ estimates are given in KSR01 and \citet{kra02}. Table~\ref{corotations} compiles
the results for the five analyzed galaxies and Figure~\ref{allrccomp} illustrates the location of the
corotation (circle) with its uncertainty range. Figure~\ref{diskrcrexp} compares the estimated corotation radii
to the disk scale length. Taking the $K$-band exponential disk scale length, the mean ratio of the corotation radius
and the disk scale length shows little variation among the objects: $2.71 \pm 0.43$, increasing to $3.04 \pm 0.49$
for the color corrected exponential disk scale lengths. Already from this rather small sample it is possible to
outline general trends for spiral galaxies. 

First, one can examine the spiral morphologies and how they reflect the dynamical state of the disk.
In the modal theory of spiral structure
\citep{ber89a,ber89b} the corotation resonance is typically located in the outer parts
of the optical disk, where the gas-to-star density ratio is larger.
The stellar spiral should exhibit a basic continuity across the corotation resonance and a reversal
of the properties of the spiral tracers, such as the locations of \ion{H}{2}-regions
\citep{ber93}. As is apparent from Figure~\ref{allrccomp}, these predictions are only partly
satisfied for the galaxies in our sample.
On the other hand, some of the galaxies reveal multi-arm or flocculent 
spiral properties that cannot be explained by linear, quasi-stationary modal theories.
Non-linear mechanisms or transient amplification processes might play important roles and
affect the location of the corotation resonance, too. The results from non-linear
orbital modelling of spiral structures \citep{pat91} indicate
that the strong, symmetric, logarithmic stellar spirals already end at the inner 4:1
resonance, located well inside the corotation resonance. In their approach,
the end of the strong spiral pattern is found at a radial range of 2 -- 3 exponential disk
scale lengths, which is compatible with a corotation radius of $\gtrsim 3\,R_{\rm exp}$.
As seen from Figure~\ref{diskrcrexp}, the corotation radii from our
galaxy sample tend to fall in that range.

Figure~\ref{allrccomp} compares these corotation radii to the
image morphologies. Clearly, the dominant two-arm pattern in these galaxies lies mostly within
the corotation resonance. This finding does not disagree with the conclusions of \citet{pat91} and
\citet{pat99}, derived from non-linear orbital modelling and N-body simulations respectively.
However, the linear modal theory of spiral structure also predicts that the corotation radius
falls into the radial range, which was found for the sample galaxies discussed here. \citet{ber93} states
$R_{\rm CR} \approx 3-4\,R_{\rm exp}$. Thus, the locations of the corotation
resonances in the sample galaxies are mostly in agreement with the predictions from linear as well
as non-linear theories. The good agreement between the measured disk properties and the theoretical
expectations from spiral density wave theories can be regarded as support of these theories.

Determining the location of the corotation resonance from hydrodynamic modelling is a rather elaborate
and time consuming process. However, looking at Figure~\ref{allrccomp} it becomes clear that 
there is no obvious common feature in the galactic disks that would enable an easy identification
of the corotation resonance otherwise. It is not
always the case that at corotation the star forming rate is largely reduced. Furthermore, there is no
characteristic feature in the amplitudes of the low Fourier components at the radius of the corotation
resonance. This lack of any reliable corotation tracer complicates the application of density wave
theories to real galaxies. Ultimately, the appearance of any real
galaxy in the universe is the product of a combination of many processes. 
In some cases, the linear modal density wave theory can be applied to describe the relevant issues,
but in many galaxies non-linear effects in the density wave dynamics and other processes seem to play a role.
Nevertheless, assuming a fixed spiral pattern rotation speed seems, to first order, to be a viable
assumption for the simulations of gas flows in fixed galaxy potentials. 
In light of this, it is a very promising conclusion that the hydrodynamic gas simulations provide
a powerful tool to learn about disk dynamics and to determine the location of resonances in spiral galaxies.

\subsubsection{Quantitative match of the 2D gas kinematics}
\begin{figure}
\begin{center}
 \resizebox{\hsize}{!}{\includegraphics{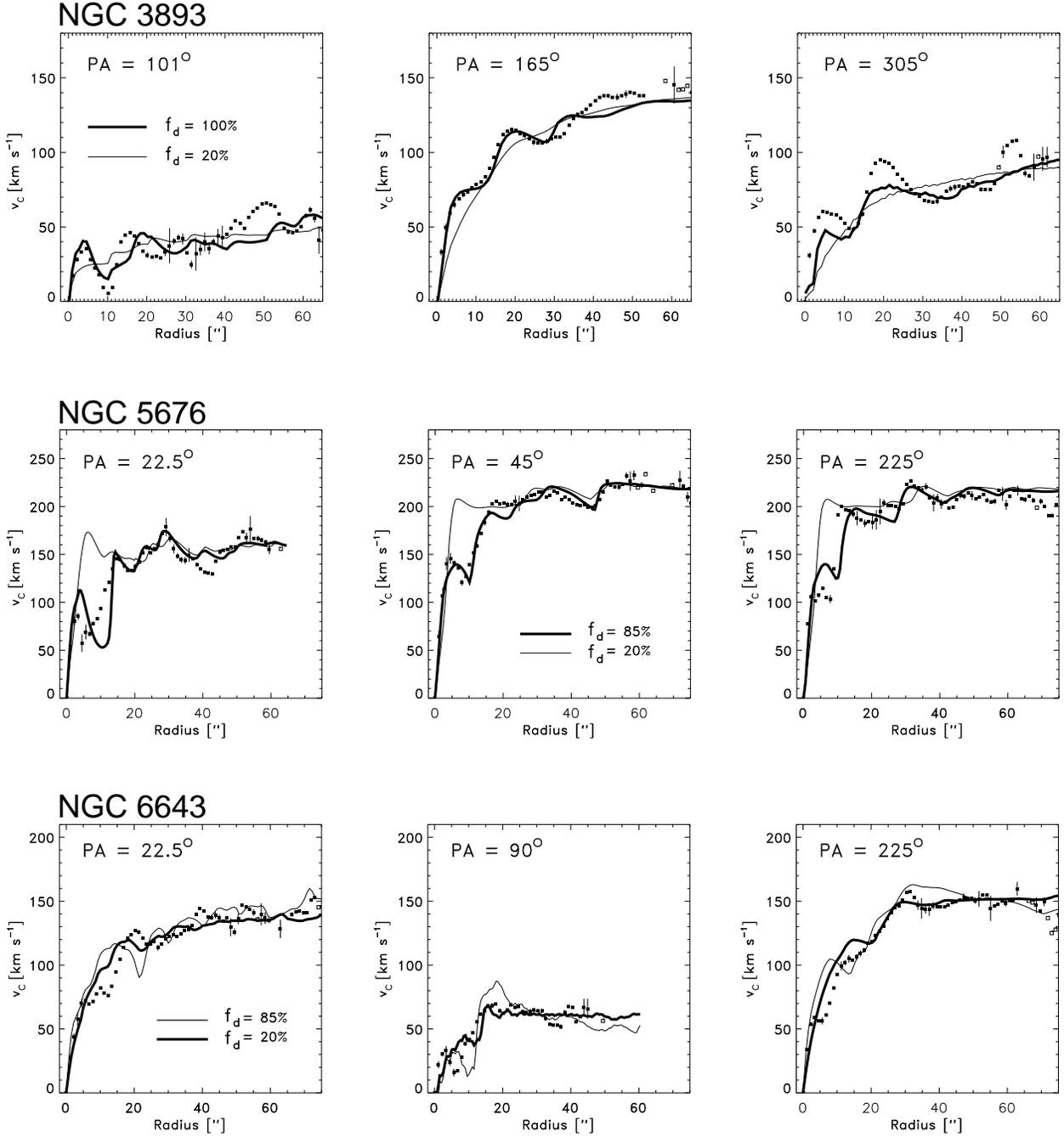}}
 \caption{Examples of the comparison between simulation results and observed kinematics (data points)
   for the galaxies NGC 3893, NGC 5676, and NGC 6643. Displayed are three position angles (PA) for each of the
   galaxies with velocity fields from ``maximal disk'' and ``minimal disk'' simulations overlaid. The best
   fitting model is indicated by the thick line.
    \label{velcuts}}
\end{center}
\end{figure}

In addition to the good agreement of the simulated gas density distribution with the disk morphology,
the modelled gas velocity fields provide a good overall match to the measured kinematics too.
This comparison can be made along the 14 or 16 position angles along which we have long-slit H$\alpha$ velocities.
Figure \ref{velcuts}\ gives an example of how good the simulations reproduce individual features in the
observed kinematics. Displayed are the rotation curves along three slit positions across the disks of the galaxies
NGC 3893, NGC 5676, and NGC 6643. Overlaid are the simulation results from each a maximum disk and a minimum
disk model with the best fitting model indicated by the thick line.
In all cases, the global structure of the velocity field could be reproduced with satisfying accuracy by the model
velocity field. This is a non-trivial result which speaks highly for the quality of the
hydrodynamics code.

In order to compare the amplitude of individual wiggles and rotation curve features to the observations,
the code was required not only to reproduce global gas dynamics but also to properly model the small scale
features. In general, this is also quite well achieved by the simulations, as it can be read from Figure
\ref{velcuts}. \citet{sly02b} discuss the details of how the positions and amplitudes of
features in the simulations depend on our modelling assumptions.
However, within the galactic disks, there are processes other than local gravity
fluctuations which create additional gas-dynamic effects. The gas dynamic traces of expanding supernova
gas shells or outflows from young star forming regions are superimposed on the larger scale
response to the gravitational potential wells and may eventually cover up the features that are
relevant for this project. In order to minimize the influence of these non-gravitational effects, we
used the hydrodynamic simulations -- for which the only driving force is gravity -- as a reference to discard
a fraction of 30 - 40\,\% from the observed data points that deviate largely from \emph{all} simulated cases as
non-gravitationally induced wiggles. This procedure is explained in more detail in \S 4.3.2.1. of \citet{kra02}.
It should be added that with this treatment only local parts of the rotation curves, such as individual wiggles
or features near the center, where small bars could play a role, were excluded and that the
global coverage of the disk is not sacrificed. Ultimately, it would be desirable to use high sensitivity two
dimensional velocity maps to provide a better understanding of the dynamical processes in the disk and to
completely eliminate this concern.

\begin{figure}
\begin{center}
 \resizebox{\hsize}{!}{\includegraphics{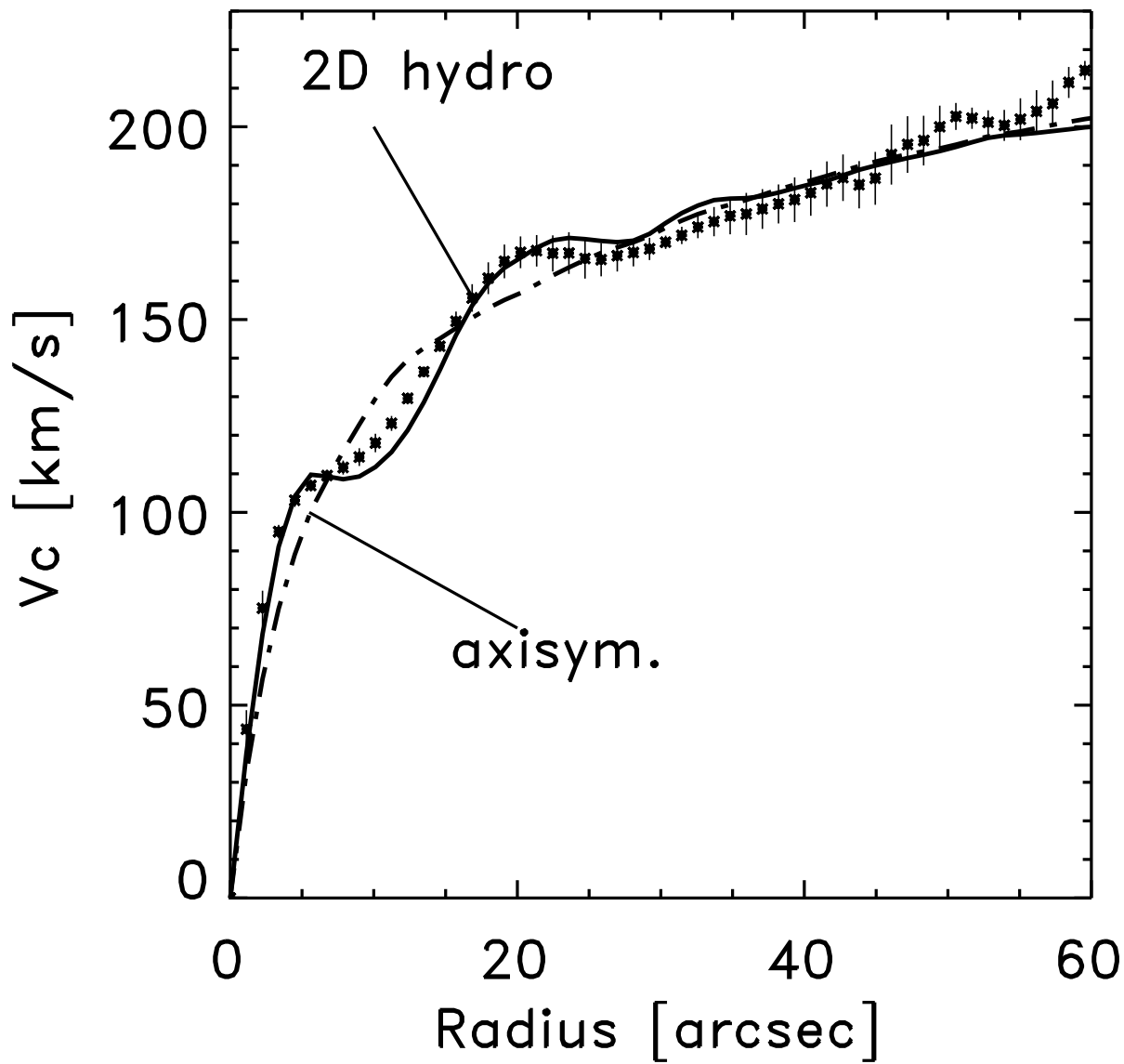}}
 \caption{Effect of the two dimensional simulation on the rotation curve (averaged around the major axis),
    displayed for NGC 3893.
    The two dimensional fit (continuous line) provides a better representation of the inner rotation curve
    than the axisymmetric fit (dash-dotted line) does.
    \label{n3893gasvel}}
\end{center}
\end{figure}

Using this technique we find that the agreement between the simulated two dimensional velocity fields and the
observed rotation curves is very good also on a wiggle-by-wiggle basis. As an example, this is shown for the
\emph{averaged} rotation curve of NGC 3893 in Figure~\ref{n3893gasvel}. About 30\,\% of the data points have
been discarded. Although the match of the large wiggle at about 12\arcsec\ radius
is not perfect, the simulated velocity field provides a better general representation of the
rotation curve features than any axisymmetric rotation curve model could do.

\subsection{Ratio of stellar to dark mass}

\begin{figure}
\begin{center}
 \resizebox{\hsize}{!}{\includegraphics{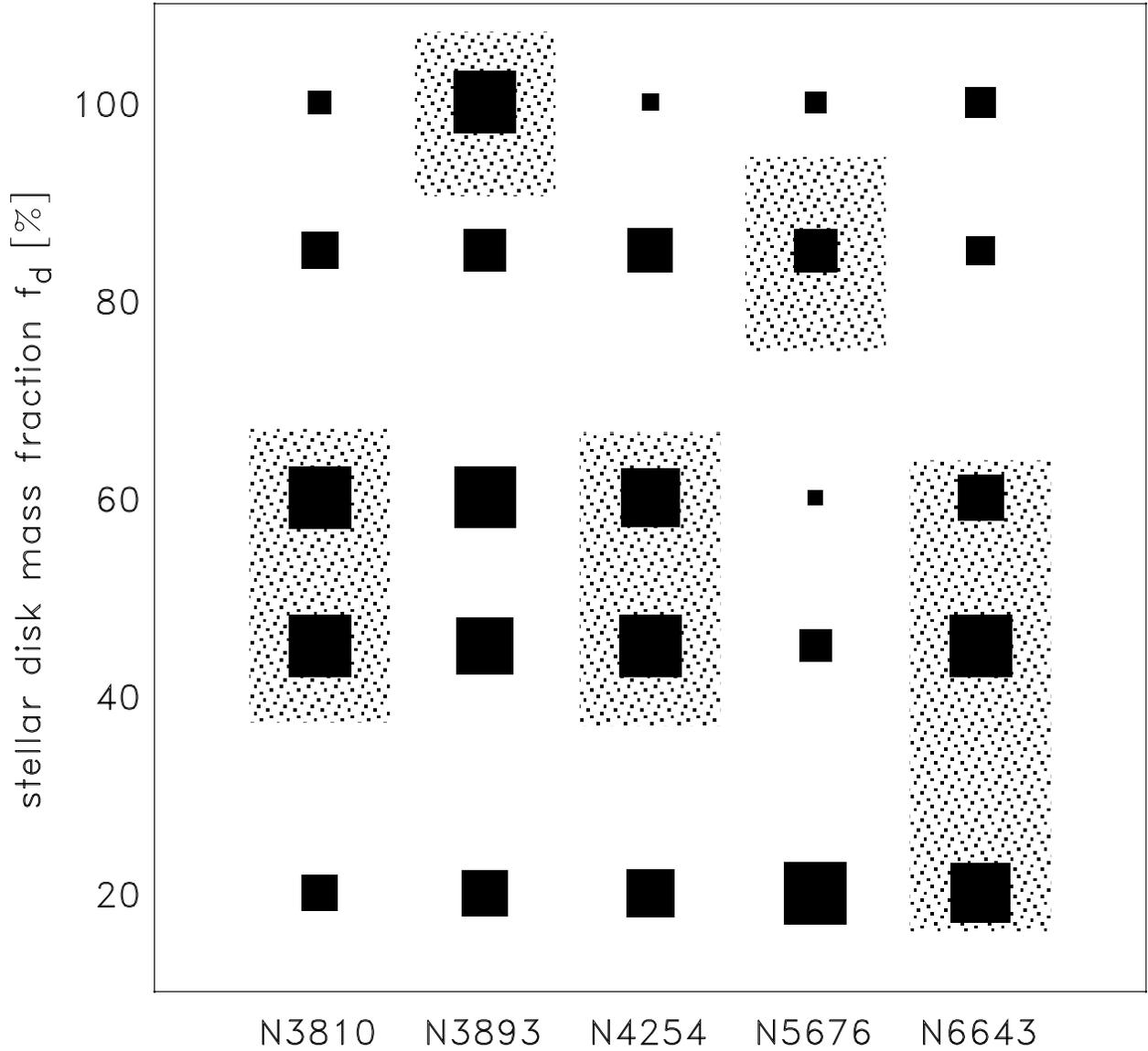}}
 \caption{Results from the least squares analysis. Larger symbols represent better $\chi^2$-values.
 The symbol sizes are scaled to give equal sizes for the best fitting points. Regions of best agreement
 between simulations and observations are marked by the dotted regions. Note that these shaded areas
 also account for a visual inspection of the fit quality.
    \label{allgalcompare}}
\end{center}
\end{figure}

Our main objective was to find the relative stellar disk to dark halo mass fractions
from the comparison of modelled gas flows within realistic galaxy gravitational
potentials of differing dark and luminous matter composition to observations. 
The results from the model-to-data comparison is presented in Figure~\ref{allgalcompare} for the five
galaxies NGC 3810, NGC 3893, NGC 4254, NGC 5676 and NGC 6643 drawing on a least squares comparison of predicted
velocity fields for differing ${\rm f_d}$ to the observed. The symbol sizes represent the probability 
-- derived from the least squares analysis -- that a certain mass model applies to the specific galaxy. The
most probable values for the disk mass fraction is indicated by the dotted regions in the plot. Since in some cases
a purely formal least squares analysis does not give an unambiguous indication of the mass composition of the galaxy, we
have to reexamine the quality of the fit by eye in order to assign a certain disk mass fraction to the specific
galaxy. In the cases of NGC 3893 and NGC 5676 this approach supported higher disk mass fraction
models \citep{kra02}. As can be seen from Figure~\ref{allgalcompare}, the precision with
which a certain scenario can be attributed to a particular galaxy is not equally good for all sample galaxies.
For NGC 3893 the disk mass fraction could be estimated quite well, placing this galaxy into the realm of
maximum disks. For NGC 3810 and NGC 6643 the least squares comparison mostly argued for a light
to medium disk scenario. The analysis for NGC 6643 was especially difficult since this galaxy exhibits only a
weak stellar spiral density wave, leading to a larger uncertainty range for the disk mass fraction.
The analysis of NGC 4254 suffered from the high abundance of strong non-gravitationally induced gas
kinematic features in the observed velocity field, which is apparent from the high abundance of small scale wiggles in
the obseved rotation curves. The main conclusion for NGC 4254 was to exclude the
strictly maximal disk solution. For NGC 5676 the $\chi^2$-analysis mostly argues for a heavy disk scenario, although
some simulations suffered from premature terminations. However, as discussed in \citet{kra02}, the simulations
proceeded to a point where the spiral pattern seemed to have reached a quasi-stationary state, making us 
confident that the conclusions for this galaxy are quite reliable.

As defined by \citet{sac97}, the designation ``maximal disk'' applies to a galactic disk when $85 \pm 10$\,\%
of the total rotational support of a galaxy at a radius R = 2.2 R$_{\rm exp}$ is contributed by the stellar
disk mass component. In the extreme case of $v_{\rm disk} = 0.75\,v_{\rm tot}$, this definition allows a
relatively massive halo to almost comprise the same amount of mass as the disk within that radius.

The above definition translates into the notation used here in the following way. The rotational support
from the stellar disk is
$v_{\rm disk}(R_{2.2}) = \sqrt{\rm f_d} v_{\star}(R_{2.2})$, where $v_{\star}(R_{2.2})$ is the rotational
support from the maximal stellar disk at a radius of 2.2 disk scale lengths. A maximal disk model also
includes a halo contribution $v_{\star}(R_{2.2}) = {\rm f_{max}}\,v_{\rm tot}(R_{2.2})$, with ${\rm f_{max}}$
typically 0.9. Accordingly, the rotational support from the disk is about 0.9 (${\rm f_d} = 100$\,\%),
0.83 (${\rm f_d} = 85$\,\%), 0.7 (${\rm f_d} = 60$\,\%), 0.6 (${\rm f_d} = 45$\,\%), and 0.4 (${\rm f_d} = 20$\,\%).
All of the ${\rm f_d}$ = 85\,\% models clearly represent maximal disk scenarios. The ${\rm f_d}$ = 60\,\%
models are at the lower border of the maximal disk realm, representing compositions with about an equal
amount of dark and luminous mass within 2 -- 3 exponential disk scale lengths. 
In light of this, Figure~\ref{allgalcompare} shows that a maximal disk scenario applies to NGC 3893 and
NGC 5676. For the other galaxies the results from the modelling favor less massive stellar disks.
NGC 3810, NGC 4254 and NGC 6643 possess stellar disks that appear to balance the mass of the dark
halo inside the optical disk (${\rm f_d} \approx 60$\,\%), although for NGC 6643 an even lighter stellar
mass component could not be excluded.

Table~\ref{allml} gives an overview of the stellar mass-to-light ratios that were derived from
the calculations of the potentials. The values are given for a maximum disk and mostly fall in the range
between M/L$_K$ = 0.6 -- 0.7. A smaller value for M/L$_K$ was only found for NGC 4254.
These stellar mass-to-light ratios agree very well with those emerging from the color correction
procedure of \citet{bel01}, which are based on population synthesis models\footnote{\citet{bel01}
used an initial mass function that was scaled to a maximal disk scenario, which was established from
measured disk kinematics. In light of this, the  population synthesis M/Ls also ultimately rely on disk dynamics.}.
Using these 
mass-to-light ratios, the total mass of the stellar disk can be inferred. The values given in Table~\ref{allml}
apply to a disk mass fraction which, as taken from Figure~\ref{vc2fdcomp},
is a reasonable assumption for the particular galaxy. These disk masses appear realistic and
scale well with the maximal velocities in the rotation curves, i.e.~attributing a high disk mass to
NGC 5676 and a light disk to NGC 3810. The mismatch for NGC 4254 most likely originates from
residual inaccuracies in the determination of the galaxy's distance, inclination and possible
calibration errors due to non-photometric conditions during the data acquisition.
The other results for NGC 4254 are only insignificantly affected by possible inaccuracies of these parameters.

\section{DISCUSSION}

\begin{figure}
\begin{center}
 \resizebox{12.8cm}{!}{\includegraphics{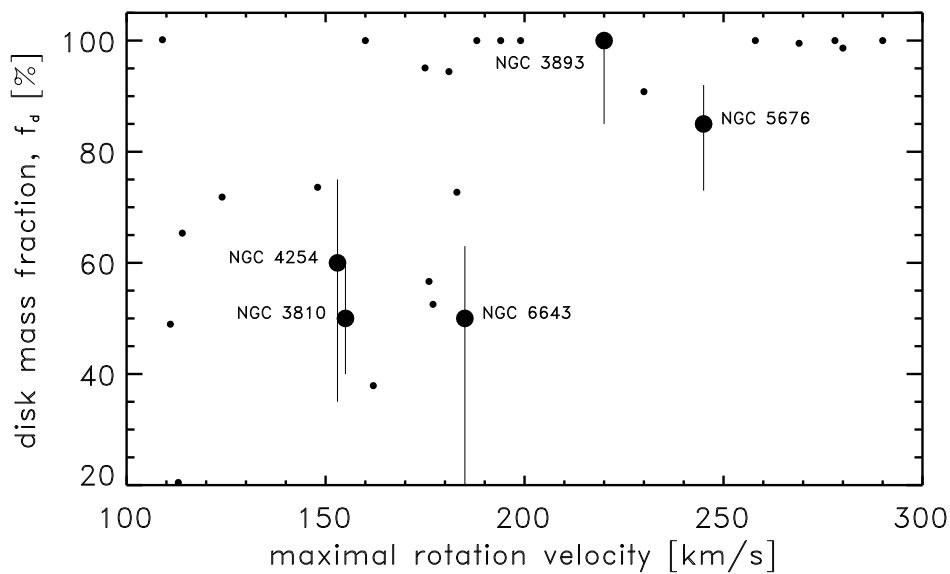}}
 \caption{The dependence of the disk mass of galaxies on their total mass, as inferred from
the maximal rotation velocity. The disk mass fraction ${\rm f_d}$ is defined in Equation~(\ref{eqn2}).
In addition to the results from this study (filled dots) results from
the study of \citet{ath87} (small dots) are included in the plot. The galaxies 
can be separated into two classes. The most massive galaxies tend to
possess disks with high mass fractions, ${\rm f_d}$. Less massive systems may exhibit
a considerable dark mass fraction.\label{vc2fdcomp}}
\end{center}
\end{figure}

We have presented a new, direct and independent estimate of the stellar mass fraction in a small sample of spiral
galaxies. With respect to ratio of luminous and dark matter in their inner parts, the population of
high surface brightness spiral galaxies does not seem to comprise an entirely homogeneous class of objects. Combining
the maximum rotation velocity, presented in Table~\ref{properties}, with the estimates for the galaxies' most
probable fractions of the stellar disk mass, reveals an intriguing trend: the most massive of the analyzed
galaxies, NGC 3893 and NGC 5676, tend to also possess the most massive stellar disks which dominate the
dynamics of the central regions. The other galaxies from the sample are less massive systems that exhibit a
maximum rotation velocity $v_c$\,$<$\,200\,km\,s$^{-1}$. For those objects the total mass of the dark halo within
the optical radius is higher and was found to, at least, equal the stellar mass. This trend is graphically
presented in Figure~\ref{vc2fdcomp}.

Our results can be compared to \citet{ath87}.
These authors inferred disk and halo mass fractions from spiral structure constraints. The small points displayed
in Figure~\ref{vc2fdcomp} represent a subsample of the total sample (given in Athanassoula et al.~1987, Table A2),
selected for late type galaxies with a bulge mass $M_{\rm bulge} < 20$\,\% $M_{\rm disk}$ and a maximum rotation
velocity of more than 100\,km\,s$^{-1}$. For comparison
the stated disk-to-halo mass ratio was converted into our scheme of quantifying the disk mass fraction ${\rm f_d}$.

It can be clearly seen that the trend that emerged from our small dataset is supported by the results from
\citet{ath87}. In this sample there is no late type spiral galaxy with a maximum rotation velocity
of over 200\,km\,s$^{-1}$ that seems to be in agreement with a substantially submaximal stellar disk. For less
massive systems this is not the case anymore. Apparently galaxies with a wide range of dark halo and disk mass
decompositions can exist and maintain spiral structure.

In light of this, the results from \citet{CR99} might be explained that in a statistical
sample of spiral galaxies the heavy and clearly maximum disk galaxies could account for only a relatively
small fraction
and most of the ``lighter'' high surface brightness galaxies exhibit already a considerable dark mass
fraction. Indeed, in the Courteau \& Rix sample, only roughly a third of the galaxies exhibit maximal rotation
velocities $v_c$\,$>$\,200\,km\,s$^{-1}$. 

However, beyond the discussion of the disk mass fraction in spiral galaxies, at present there is growing evidence
that galaxy size dark halos are not cuspy in their centers, as claimed by standard cold dark matter (CDM)
models. \citep{moo94,flo94,moo99b,sal00,bor01,deb01,vdb01}. This circumstance was first
found for low surface brightness and dwarf galaxies, that supposedly are comprised of a massive dark halo.
Most of the authors argued that this also applies to high surface brightness galaxies. For the present
analysis, the functional form of the dark halo profile has only a second order effect on the results
as long as the models describe well the overall shape of the observed rotation curves.
Thus, we decided not to distinguish between different dark halo profiles. However, the results
from this study, which mostly confirm the strong dynamic influence of the stellar disk in massive high surface
brightness spiral galaxies within the optical radius, provide very little room for a strongly
cusped halo core.

\section{SUMMARY}

With this project it has been demonstrated that an analysis similar to the one applied to barred galaxies
\citep{eng99,wei01} can also be applied to non-barred spiral galaxies. This application is on one hand
more difficult and problematic, since the spiral arms provide a much less prominent
dynamical feature in the velocity field than a strong bar does. On the other hand, for pure disk galaxies
the vertical scale height is smaller in the central regions than in strong bars that often show peanut
shapes when viewed edge-on. This fact reduces systematic uncertainties that arise from motions along the
z-axis which are superimposed onto the circular motions that are studied. The main points that could be
learned from this study can be summarized in the following way:

\begin{itemize}
\item The color correction, described in Section~\ref{coco}, allows us to derive the stellar mass density from
 any multi color imaging data. As a correction to the K-band photometry Equation~\ref{equation} has proven to
 work very well. We encourage its use for all studies that depend on the mass distribution in galaxy disks.
\item The BGK hydrodynamics code, applied to the problem \citep{sly02b}, is a powerful tool and it can be
 used successfully to simulate the interstellar gas in the disks of galaxies. The parameters that dominate
 the morphology and the kinematics of the gas simulations are the non-axisymmetric stellar mass component
 and the pattern speed of the spiral structure.
\item Adopting a globally fixed pattern speed for the rotation of the spiral seems to be a viable approximation
 for the simulation process. Moreover, from analyzing the morphology of the simulated gas density distribution 
 it could be found for the sample of five galaxies that the corotation radius $R_{\rm CR} = 2.71 \pm 0.43 R_{\rm exp}$,
 increasing to $3.04 \pm 0.49 R_{\rm exp}$ for the color corrected exponential scale lengths representing the stellar disk mass.
 In all five cases the strong stellar spiral ends inside the corotation resonance.
\item The detailed comparison of the velocity fields is challenging, but it is possible to put constraints on the relative mass
 fractions of the inner disk. The results from this study indicate that high surface brightness galaxies possess
 maximal disks if their maximal rotation velocity $v_c$\,$>$\,200\,km\,s$^{-1}$. If the maximal rotation velocity is
 less, the galaxies appear to have submaximal disks.
\end{itemize}

\acknowledgements
We would like to thank the Calar Alto staff for their support in obtaining the data for this study.
AS acknowledges the support of a fellowship from the UK Astrophysical Fluids Facility (UKAFF).

\clearpage

\clearpage

\clearpage

\clearpage

\clearpage

\clearpage

\clearpage

\clearpage

\begin{deluxetable}{lcccccccc}
\tablewidth{0pt}
\tablecaption{Properties of the sample galaxies. \label{properties}}
\tablehead{ \colhead{Galaxy} & \colhead{$B_{\rm tot}$} & \colhead{K\arcmin$_{\rm tot}$} & \colhead{PA} & \colhead{$i$} & \colhead{max $v_{\rm rot}$} & \colhead{dist.} & \colhead{optical} & \colhead{reference} \\
 & [mag] & [mag] & [\degr] & [\degr] & [km\,s$^{-1}$] & [Mpc] & image & }
\startdata
NGC 3810 & 11.4 & 8.02 & 22 & 46 & 155 & 13.5 & B & Frei et al. 1996 \\
NGC 3893 & 11.23 & 7.94 & 166 & 42 & 220 & 17 & B & Tully et al. 1996 \\
NGC 4254 & 10.2 & 6.85 & 67.5 & 41.2 & 153 & 20 & g & Frei et al. 1996 \\
NGC 5676 & 11.7 & 8.03 & 46.5 & 63.2 & 245 & 33 & V & H{\'e}raudeau \& Simien 1996 \\
NGC 6643 & 11.8 & 8.18 & 39 & 57.5 & 185 & 23 & B & H{\'e}raudeau \& Simien 1996 \\
\enddata
\end{deluxetable}

\clearpage

\begin{deluxetable}{cccc}
\tablewidth{0pt}
\tablecaption{Corotation radii and exponential scale lengths for the galaxies. \label{corotations}}
\tablehead{ \colhead{Galaxy} & \colhead{$R_{\rm CR}$} & \colhead{$R_{\rm exp}$($K$\arcmin)} & \colhead{$R_{\rm exp}$($K$\arcmin$_{\rm corr}$)} \\
 & [kpc] & [kpc] & [kpc]}
\startdata
NGC 3810 & 3.15 $\pm$ 0.5 & 1.07 & 0.913 \\
NGC 3893 & 5.5  $\pm$ 0.5 & 1.80 & 1.74 \\
NGC 4254 & 7.5  $\pm$ 1.1 & 3.54 & 3.06 \\
NGC 5676 & 11.0 $^{+ 1.0} _{- 2.0}$ & 3.59 & 3.12 \\
NGC 6643 & 6.5 $^{+ 1.5} _{- 0.5}$ & 2.72 & 2.48 \\
\enddata
\tablecomments{All values are calculated for the distance given in Table~\ref{properties}.
$R_{\rm exp}$($K$\arcmin$_{\rm corr}$) specifies the color corrected
exponential scale lengths.}
\end{deluxetable}

\clearpage

\begin{deluxetable}{lccccc}
\tablewidth{0pt}
\tablecaption{$\chi^2/N$-values from the least squares analysis.\label{allchi2}}
\tablehead{ \colhead{Galaxy} & \colhead{${\rm f_d} = 20$} & \colhead{${\rm f_d} = 45$} & \colhead{${\rm f_d} = 60$} & \colhead{${\rm f_d} = 85$} & \colhead{${\rm f_d} = 100$}}
\startdata
NGC 3810 & 0.800 & 0.550 & 0.555 & 0.787 & 1.069 \\
NGC 3893 & 1.775 & 1.555 & 1.480 & 1.897 & 1.451 \\
NGC 4254 & 1.146 & 0.959 & 1.007 & 1.206 & 2.303 \\
NGC 5676 & 0.706 & 1.109 & 1.817 & 0.908 & 1.463 \\
NGC 6643 & 1.400 & 1.349 & 1.651 & 2.284 & 2.183 \\
\enddata
\end{deluxetable}

\clearpage

\begin{deluxetable}{lcccccc}
\tablewidth{0pt}
\tablecaption{Maximum disk stellar $K$-band mass-to-light ratios and the associated stellar disk mass. \label{allml}}
\tablehead{ \colhead{Galaxy} & \colhead{(M/L$_K)_{\rm max}$} & \colhead{(M/L)$_{\rm color}$} & \colhead{M$_{\rm disk\,|\,max}$}& \colhead{M$_{\rm disk\,|\,f_d}$}}
\startdata
NGC 3810 & 0.63 & 0.55 & $1.48 \times 10^{10}$ & $0.89 \times 10^{10}$ \\
NGC 3893 & 0.56 & 0.54 & $2.32 \times 10^{10}$ & $2.32 \times 10^{10}$ \\
NGC 4254 & 0.23 & 0.74 & $3.53 \times 10^{10}$ & $2.12 \times 10^{10}$ \\
NGC 5676 & 0.67 & 0.83 & $9.55 \times 10^{10}$ & $8.12 \times 10^{10}$ \\
NGC 6643 & 0.71 & 0.65 & $4.23 \times 10^{10}$ & $2.54 \times 10^{10}$ \\
\enddata 
\tablecomments{Given are the maximal disk $K$-band M/Ls. (M/L$_K)_{\rm max}$ corresponds to a stellar mass fraction
${\rm f_d = 100}$; (M/L)$_{\rm color}$ were derived using the relations by Bell \& de Jong (2001) for the
respective overall galaxy colors. The disk mass M$_{\rm disk\,|\,f_d}$ has been scaled for the galaxies' most probable disk
mass fraction ${\rm f_d}$. All values are given in solar units.}
\end{deluxetable}

\end{document}